\documentclass{article}

\usepackage{fontspec}

\usepackage[preprint,nonatbib]{neurips_2026}
\usepackage[sort]{natbib}
\usepackage{xurl}
\usepackage[utf8]{inputenc} 
\usepackage[T1]{fontenc}    
\usepackage{hyperref}       
\usepackage{url}            
\usepackage{booktabs}       
\usepackage{amsfonts}       
\usepackage{nicefrac}       
\usepackage{microtype}      
\usepackage{xcolor}         
\usepackage[most]{tcolorbox}
\usepackage{listings}

\usepackage{amsmath}
\usepackage{graphicx}

\newfontface\kumizraw{KumizRegular.ttf}

\newfontface\kumizlessspace[
    LetterSpace=-2.5
]{KumizRegular.ttf}

\newcommand{\xethintext}[2]{%
  #2%
}

\newcommand{\kumiz}[1]{%
  {\color{black}%
   \kumizraw
   \xethintext{0.2}{#1}%
  }%
}

\newcommand{\kumizthin}[1]{%
  {\color{black}%
   \kumizlessspace
   \xethintext{0.65}{#1}%
  }%
}

\usepackage{cleveref}
\crefname{figure}{Fig.}{Figs}
\crefname{table}{Tab.}{Tabs}
\crefname{section}{Sec.}{Secs}
\Crefname{figure}{Fig.}{Figs}
\Crefname{table}{Tab.}{Tabs}
\Crefname{section}{Sec.}{Secs}

\usepackage{pdfrender}
\newcommand{\methodbold}{{\kumiz{Scrub Data}}}
\newcommand{\method}{{\kumizthin{Scrub Data}}}

\title{\methodbold{}: A Framework for Reproducible\\ Data Curation with AI Coding Agents
}

\author{
  Justin Kay\\MIT\\\texttt{kayj@mit.edu}
    \And
    Shir Bar\\MIT\\\texttt{shir3bar@mit.edu}
    \And
    Sara Beery\\MIT\\\texttt{beery@mit.edu}
}

\begin{document}

\maketitle

\vspace{-8pt}
\begin{figure}[ht]
    \centering
    \includegraphics[width=\linewidth]{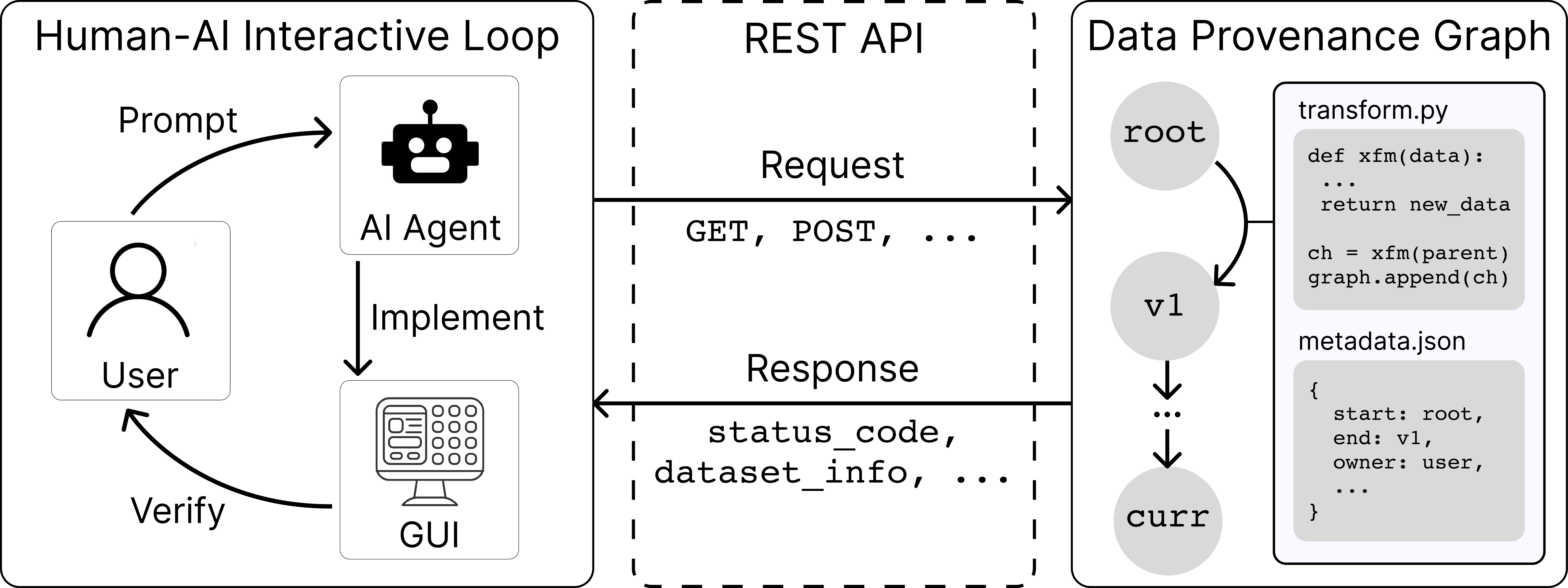}
    \vspace{-16pt}
    \caption{\small\textbf{The \methodbold{} framework for agent-assisted data curation.} Users interact with an AI agent to iteratively execute data curation operations---such as cleaning and merging raw datasets---and develop graphical interfaces to validate their changes (left). Dataset modifications are recorded in an executable data provenance graph, maintaining the full version history in a reproducible manner (right). These components are tied together by a lightweight web framework and REST API that separates agents and raw data, ensuring that the data graph remains well-formed while providing easy access patterns to frontend interfaces (center).}
    \label{fig:framework}
\end{figure}

\vspace{-3pt}
\begin{abstract}
\vspace{-5pt}
Data curation---the process of collecting, cleaning, and joining raw datasets for unified analysis---is a time-consuming yet crucial part of any data science project.
With the emergence of powerful agentic coding tools, it is tempting to ``vibe curate’’---\textit{i.e.}, to prompt AI agents to perform data curation operations and blindly trust their outputs in order to finish work more quickly.
However, this can exacerbate two key challenges that already plague manual data curation workflows:  \mbox{\textbf{reproducibility}}---the ability to trace the exact set of modifications applied to raw datasets---and \textbf{verifiability}---the ability to audit changes and confirm that data is processed properly.
To address these issues, we introduce \methodbold{}, a framework that enables the use of AI agents in data curation workflows while transparently maintaining both reproducibility and verifiability. 
The framework involves an interactive data curation loop between a user and an AI agent, backed by a data provenance graph that tracks and versions all dataset updates. The data graph requires that all updates are formatted as self-contained executable steps, allowing any version of a dataset to be reconstructed from raw data by playing forward the transformations stored in the graph.
This workflow takes place within a lightweight web application that is designed to be modified by users and agents in real-time to create custom visualization tools for identifying data curation needs and verifying outcomes.
We demonstrate the usefulness of the framework with a real data curation use case from animal movement ecology. Starting with 89M raw GPS coordinates, we use the 
framework to curate a benchmark dataset of 2.5M coordinates to be used for machine learning or statistical analysis. 
The framework is available for use and extension at \url{https://github.com/justinkay/scrubdata}.
\vspace{-10pt}
\end{abstract}

\section{Introduction}

Data curation---the time-intensive process of collecting, cleaning, and joining raw data sources to a common analysis-ready format---is an important part of any data science pipeline. 
Careful curation allows data to be interpreted consistently and used reliably in downstream analyses~\cite{chapman2005principles,wickham2014tidy,dasu2003exploratory,chu2016data,rahm2000data,kandel2011research,stonebraker2013data}.
For example, in ecology, data curation is often necessary to join data collected from heterogeneous sources, such as different camera or sensor hardware with different output formats or datasets stored using different metadata standards, and to consistently address sensor-specific noise or other errors encountered during data collection ~\cite{chapman2005principles,zizka2020no,ribeiro2022bdc,hurley2022addressing,brousil2023improving}. 
Common issues include missing or ambiguous values, duplicate records, inconsistent units or formats, erroneous geographic coordinates, and other domain-specific inconsistencies or anomalies.
If data are to be used for statistical or machine learning-based analysis, curation also involves preliminary experimental design, such as creating representative held-out sets of data for evaluation or testing, while verifying that there is no leakage between training/fitting and testing/evaluation data~\cite{kapoor_leakage_2023}. 

Data curation workflows are complex, 
often comprising numerous modifications to underlying raw data, with different data sources requiring different processing. 
Due to the heterogeneity of real-world datasets, much of this work is bespoke, requiring custom code for both implementing dataset transformations and verifying transformations through relevant exploratory analysis~\cite{stonebraker2013data,dasu2003exploratory,dutoit2012graphical,epperson2023dead}.

\begin{figure}
    \centering
    \includegraphics[width=\linewidth]{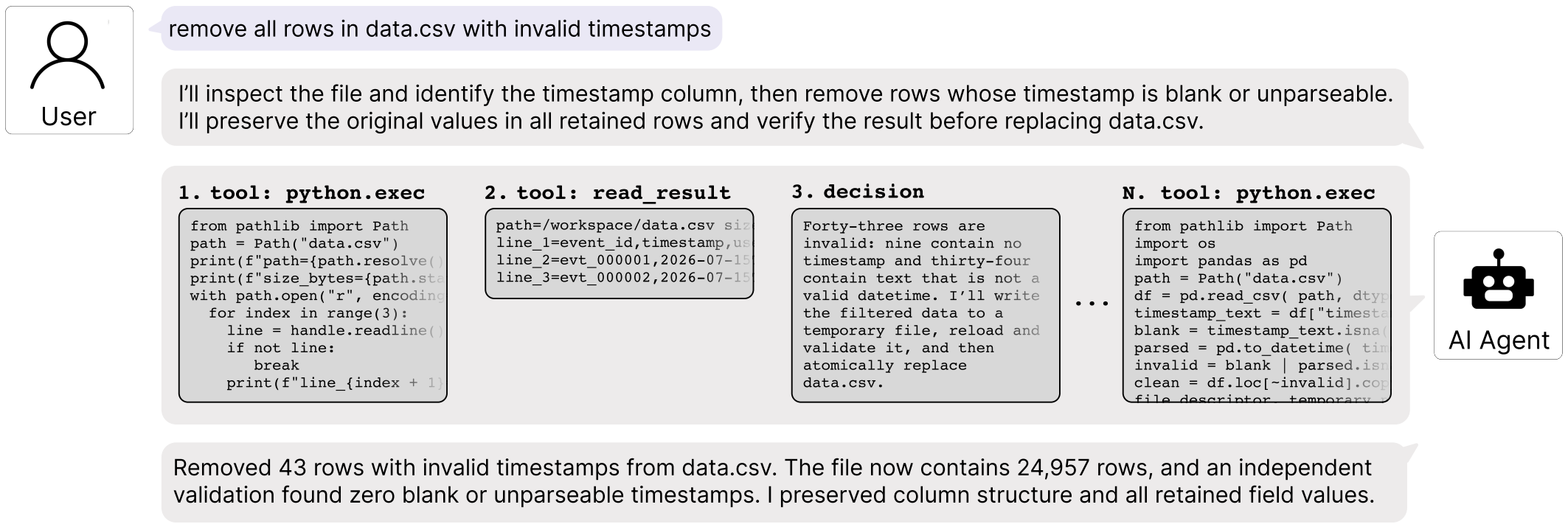}
    \caption{\small\textbf{Example AI agent trajectory during data curation work.} A user interacts in natural language with an LLM via an agent framework. In response to a request for a data cleaning operation (in this case, removing rows from a CSV file with invalid timestamp entries), the agent performs a sequence of operations constituting a \textit{trajectory}. In addition to reading and writing text output, what distinguishes agents from standard LLMs is their ability to make \textit{tool calls}: for example, they may execute scripts or shell commands, which enable many other operations such as reading and writing files. In this example, the agent executes a sequence of Python scripts to read and inspect the user's CSV file to understand where in the file the timestamps are stored and how they are formatted; modify the file as requested; and save a new version to disk.
    \textbf{While powerful, without additional structure such a workflow poses risks for both reproducibility and verifiability of data modifications}. 
    }
    \label{fig:agentconvo}
\end{figure}

To accelerate this work, many practitioners are turning to \textit{AI agents}---software systems that can autonomously write code and perform operations on behalf of a user in response to natural language prompts. 
An agent framework is a wrapper around a large language model (LLM) that provides it with additional \textit{tool use} capabilities---that is, the LLM is allowed to invoke external programs and interfaces, such as a command-line interface, Python interpreter, or file system, and observe their outputs. 
The agent can then use these observations to decide which action to take next.
Consider, for example, a common data cleaning operation such as removing entries from a CSV file with invalid values in some column of interest. We illustrate a sample agentic workflow where the column of interest includes timestamps in \cref{fig:agentconvo}. Instead of writing code, a user specifies their request in natural language. In response to this request, an agent might inspect the dataset, identify the relevant column, write and execute a filtering operation, validate the resulting file, and report the changes to the user.
As agentic coding systems have gotten more capable, particularly for writing code~\cite{openai_codex_2026,anthropic_claude_code_2026,jimenez2024swe}, using them for data curation in this manner is an increasingly appealing option, and can help save time, reduce bugs, and open doors for non-programmers.

However, the use of AI agents for dataset curation risks further exacerbating two key problems in data science: \textit{reproducibility} and \textit{verifiability}. 
Dataset curation workflows are already notoriously difficult to reproduce~\cite{sawchuk2021computational,miyakawa2020no,paullada2021data,gebru2021datasheets}. In an ideal world, curation operations are always recorded in scripts with exactly repeatable outcomes that can be audited later by dataset users---requiring fixed random seeds, versioned artifacts, and other code-intensive bookkeeping.
In practice, however, these operations are assembled ad hoc without rigorous documentation or data versioning, and may be spread across multiple scripts and other operations, including untraceable manual modifications to raw data files.
Because agents can act autonomously, they can perform these kinds of untraceable actions \textit{en masse},
modifying raw data files repeatedly without logging a record of their actions, and making it impossible to examine or reproduce the exact edits performed. 
If unwanted changes are made, it may be impossible to roll them back if backups of previous versions were not made. 

Further, because modern agents are so capable, it is tempting to trust their outputs without verifying them directly, a fast-and-loose approach to the use of AI agents that has been dubbed ``vibe coding'' in the software development community~\cite{sapkota2025vibe,toitechdesk2025vibecoding,tran2026vibe}. 
Even if not ``vibe curating,'' the rapid pace of agent-based work means that the time required to validate data modifications by hand quickly exceeds the time required to execute more prompts, making errors more likely to fall through the cracks.
Yet it has been shown that even minor changes to data preparation can have significant impacts on scientific conclusions drawn from the processed data~\cite{kapoor_leakage_2023,fuhrding2022influence,pawlik2019link}. 
Alternatively, one may use a second ``verifier'' agent to help check for mistakes~\cite{li2025generation,li2025process,lightman2024let}, however verifiers themselves are also imperfect~\cite{panickssery2024llm,wang2024large}, and may miss problems that require domain-specific knowledge to verify. 

To address these limitations and enable the scientific community to safely benefit from the efficiency agentic systems provide, we introduce a new framework called \method{} that enables the use of AI agents for dataset curation in a more reproducible and verifiable way.
The framework consists of a \textbf{sc}affold around existing AI agent tools that provides guardrails around their behavior and maintains \textbf{r}eprod\textbf{u}ci\textbf{b}ility of all data operations. \method{} 
consists of two components (see \cref{fig:framework}): 

\noindent
\textbf{(1)~A data provenance graph} that specifies and enforces the format of dataset modifications and robustly tracks versioning history. In the \method{} framework, agents do not modify data files directly. Instead, all data modifications are mediated through API access to a backend server that manages a graph representation of the full dataset version history. In the graph, each dataset version is a node, and edges between nodes represent standalone executable scripts that transform one dataset version to the next. This means that any version of the dataset can be reconstructed at any time starting with raw data and playing forward the executable modifications stored in the graph. During data curation, these modifications are proposed by a user, created and submitted by an agent, and then can be validated via the second key component of the framework: 

\noindent
\textbf{(2)~Human-in-the-loop validation interfaces} developed by users concurrently during data cleaning for verifying the correctness of data modifications. Agents excel at simple web application development, particularly when using standardized programming languages and software frameworks and when applications are small enough to fit in model context~\cite{tran2026vibe,vichare2025webdev}. We provide a scaffold for a simple web-based application built with FastAPI~\cite{lathkar2023high} that can be easily modified by AI agents in response to user requests for new visualization components. This enables users to create custom, interactive validation workflows that are specific to the needs of their data much faster than through scripting by hand.
User interface components can also be created to perform dataset update modifications through the API to further speed up data cleaning. 

We demonstrate the usefulness of the \method{} framework by using it to curate a large benchmark dataset for animal movement ecology. Starting from a dataset of 89M GPS points, we use the framework to clean, select, and validate a subset of 2.6M points for developing and evaluating predictive methods. We document the process that we used, from empty application to completed benchmark, as one end-to-end example of how the framework can be used in real data workflows.

\section{The \methodbold{} Framework}

\method{} addresses reproducibility and verifiability issues with agent-based data curation through two mechanisms: a data provenance graph and a customizable human-in-the-loop validation workflow. See \cref{fig:framework} for an overview of the framework. 

At a high level, the framework is implemented as a lightweight web application that is \textit{designed to be modified by AI agents} in response to user requests. This is an important design decision that is unique to agentic workflows---the software itself is malleable and is meant to be customized by each individual user to suit their needs. 
Meanwhile, the data itself is separated from agents by a predefined application programming interface (API).
Agents do not modify data directly; instead, they make requests to the API that are then verified, executed, and logged by a backend server, ensuring that all dataset modifications are standalone and reproducible operations. 
In this section we describe the structure of the data provenance graph (\cref{sec:datagraph}), as well as the application architecture and the validation workflow it supports (\cref{sec:scaffold}).

\subsection{Data provenance graph}
\label{sec:datagraph}

\textit{Data provenance} refers to the historical record of how a dataset was created and transformed over time, including the operations applied to it and the relationships between successive dataset versions. 

\method{} enables data provenance by representing the dataset history of a project as a directed acyclic graph (DAG). A graph consists of \textit{nodes} connected by \textit{edges}. In \method{}, nodes correspond to dataset versions, and edges correspond to modifications that transform one dataset version into another. ``Directed'' means that 
transformations between dataset versions are unidirectional; a directed edge moves one-way starting from a \textit{parent} node and ending at a \textit{child} node. ``Acyclic'' means that no directed path can ever return to its starting node, which would create infinite loops of modifications.

\noindent
\textbf{DAG nodes.} 
Each node in the provenance graph represents one dataset version, \textit{i.e.}, the collection of all data files associated with a project at a particular point in the data curation workflow.
The \textit{root} node is the first node in the graph and is created on initialization, containing all the raw data in the root dataset.
After initialization, the graph is append-only; previously created dataset versions are never modified by later steps.
Each subsequent child node that is appended to the graph represents a new dataset version produced by a successful transformation step applied to its parent node. We provide more details on how we do this in a storage-efficient manner in Appendix~\cref{appendix:dag}.
The framework maintains a pointer to one node in the graph, referred to as the \emph{current dataset}. The current dataset is the version shown to the user by default and is used as the default input for new transformation steps. 


\noindent
\textbf{DAG edges.} 
Each edge in the provenance graph represents a self-contained, executable data curation step that transforms a parent dataset version into a child dataset version. 
By default we support Python as the executable representation of a transformation step. To create a step, a user submits a standalone transformation script (\textit{i.e.}, a \texttt{.py} file, in most cases generated by an agent or through a custom interaction in the UI).
The framework then executes the transformation script, writes the affected output files to disk, and writes a summary of the result.
If the script exits successfully, the framework verifies that all output files have been successfully created, and then constructs a new child node.
The new child dataset version then becomes the current dataset.

\subsection{Using the framework}
\label{sec:scaffold}

The \method{} framework is distributed as a barebones codebase that provides scaffolding for users and agents to extend according to their needs. Specifically, this consists of an \textit{application scaffold}---\textit{i.e.,} a simple starter codebase---and a \textit{prompt scaffold}, a set of usage instructions for agents to ingest as context.

\textbf{Application scaffold.} The starter codebase defines a simple client-server web application built on FastAPI~\cite{lathkar2023high} consisting of a backend and a frontend. 
The backend is a Python package containing the data provenance graph execution logic. The frontend is ordinary HTML, CSS, and JavaScript. 

The frontend and backend communicate through a REST API~\cite{masse2011rest}, a collection of named HTTP routes that the browser can call in order to request or send  information to the backend. 
For example, the frontend can submit a \texttt{GET} request to obtain a list of available projects or the contents of a particular data file; or submit a \texttt{POST} request to create a transformation step or move the current-dataset pointer to another node in the graph. 
The browser does not need to know anything about how datasets are stored on disk or how the graph is represented internally; it only needs to send and receive structured JSON messages through these routes. See Appendix~\cref{appendix:api} for a full list of default API endpoints packaged with the framework.
Users, in collaboration with agents, can extend these endpoints as appropriate for their application, for instance by adding new routes to compute a project-specific dataset summary or submit a common data cleaning operation.

\textbf{Prompt scaffold.} A collection of documentation called a \textit{prompt scaffold} is included to guide agents in their interaction with the codebase~\cite{lulla2026impact}. 
This documentation provides context for the included capabilities and how the framework is intended to be used and extended. In particular, instructions are provided for how agents should format the \texttt{.py} transformation scripts submitted to the API as proposed DAG edges. 
Agents ingest this context upon initialization, so users do not need to provide general instructions on how to use the framework and can start data curation right away. 

\textbf{Intended workflow.} 
To begin working with the codebase, users: \textbf{(1)}~Copy their raw data into a \texttt{data/} directory inside the \method{} codebase, \textbf{(2)}~Launch an agent instance from their terminal from inside the codebase, for example by running  \texttt{codex}, \texttt{claude code}, or \texttt{opencode}, and \textbf{(3)}~Launch the web application from another terminal window with \texttt{python server.py}, and direct their browser to the appropriate port on \texttt{localhost}. 

The \method{} workflow proceeds as follows: 
Users \textbf{(1)}~create custom user interfaces (UIs) using agentic coding tools to explore their data interactively and identify data curation requirements, \textbf{(2)}~prompt AI agents to perform data transformation operations or to implement UI components that trigger data transformation operations, \textbf{(3)}~validate changes through existing UI components or build new UI components to validate changes, and \textbf{(4)}~repeat until data curation is complete. 
An executable record of the entire data curation process is saved alongside the dataset, allowing all modifications to be inspected and reproduced.

\section{Case study: Creating \textsc{MoveBench}}
\label{sec:casestudy}

To demonstrate how the \method{} framework can be used in practice, we summarize our own experience using it to create \textbf{\textsc{MoveBench}}, a first of its kind large-scale benchmark dataset for animal movement ecology.\footnote{Currently in submission; this preprint will be updated when the benchmark is made publicly available.}

\textsc{MoveBench} focuses on GPS trajectory data collected from wildlife---sequences of \texttt{(latitude, longitude, time)} tuples that describe the movement of individual animals over space and time. Wildlife tracking data spans an enormous range of scales and resolutions for an increasingly diverse array of species, supporting ecology research and conservation practice~\cite{kays2015terrestrial,verzuh2026aligning,nathan2008movement}.
There is growing interest in developing predictive methods for animal movement~\cite{forrest2026predicting,forrest2025predicting,potts2022assessing,signer2024simulating,cifka2023moveformer, ranc2022memory,hacker2026making}, yet no established benchmark for training and evaluating methods on a standardized movement dataset.
Assembling such a benchmark requires significant data curation work, including synthesizing heterogeneous, multi-scale, multi-resolution, and time-varying data sources. 
These challenges were a key motivation in our development of \method{}.

\textsc{MoveBench} focuses on the predictive task of \textit{probabilistic movement forecasting}: given an animal's current location and past movement history, where will it be at a specified time horizon in the future? 
During initial discussions, we established that two generalization challenges of interest would motivate our train/test splits: \textit{temporal} generalization---how well models trained on a set of individuals generalize to future movements of those same individuals---and \textit{individual} generalization---how well models trained on one set of individuals generalize to a different set of individuals of the same species.
These decisions motivated the desired outcomes of data curation described below.

While every user's experience with \method{} will be different due to differing data curation needs, our experience using it to develop \textsc{MoveBench} can be illustrative of the kinds of operations and workflows that are possible with the framework. 
The work described in the following sections took place in March 2026 using Codex and GPT-5.4.

\subsection{Raw data acquisition and ingestion} 



We started by collecting a large set of raw animal movement data from several sources. 
Animal movement data has historically been siloed across research and conservation groups around the world performing their own independent studies, and there have been limited collective efforts to centralize data. 
In the last decade, this has begun to change, most prominently with the movement data repository Movebank~\cite{web205Movebank}, which provides a hosting platform for movement data as well as a common format for standardization and interoperability. Movebank currently hosts over ten billion animal observations from over ten thousand independent studies (\cite{web205Movebank}, accessed: August 2026). A small (yet growing) number of these studies have been released with Creative Commons licenses, making them usable by outside researchers---we started with 215 such studies for our curation~\cite{movebankMovebank}. We then added an additional fourteen candidate studies sourced from a literature review of recent movement methods papers that published empirical data alongside the method, as well as five additional studies suggested by domain experts in the Movement Biodiversity Observation Network (Move BON)\cite{geobonMove} Working Group on AI. In total, this resulted in 234 studies and 89M raw GPS points that comprised our curation starting point. 


We downloaded all studies from Movebank to a \texttt{data/movebench/} directory in the \method{} app. We then set out to convert all other studies to Movebank's standard CSV format. We did so by first downloading them to a temporary directory, then working with \texttt{codex} to write custom conversion scripts for each study. 

These conversions---because they were implemented by agents---immediately provided motivation to develop initial visualization tools for validation.

\subsection{Initial visualization development}
\label{sec:initialviz}

\begin{figure}[t]
    \centering
    \includegraphics[width=\linewidth]{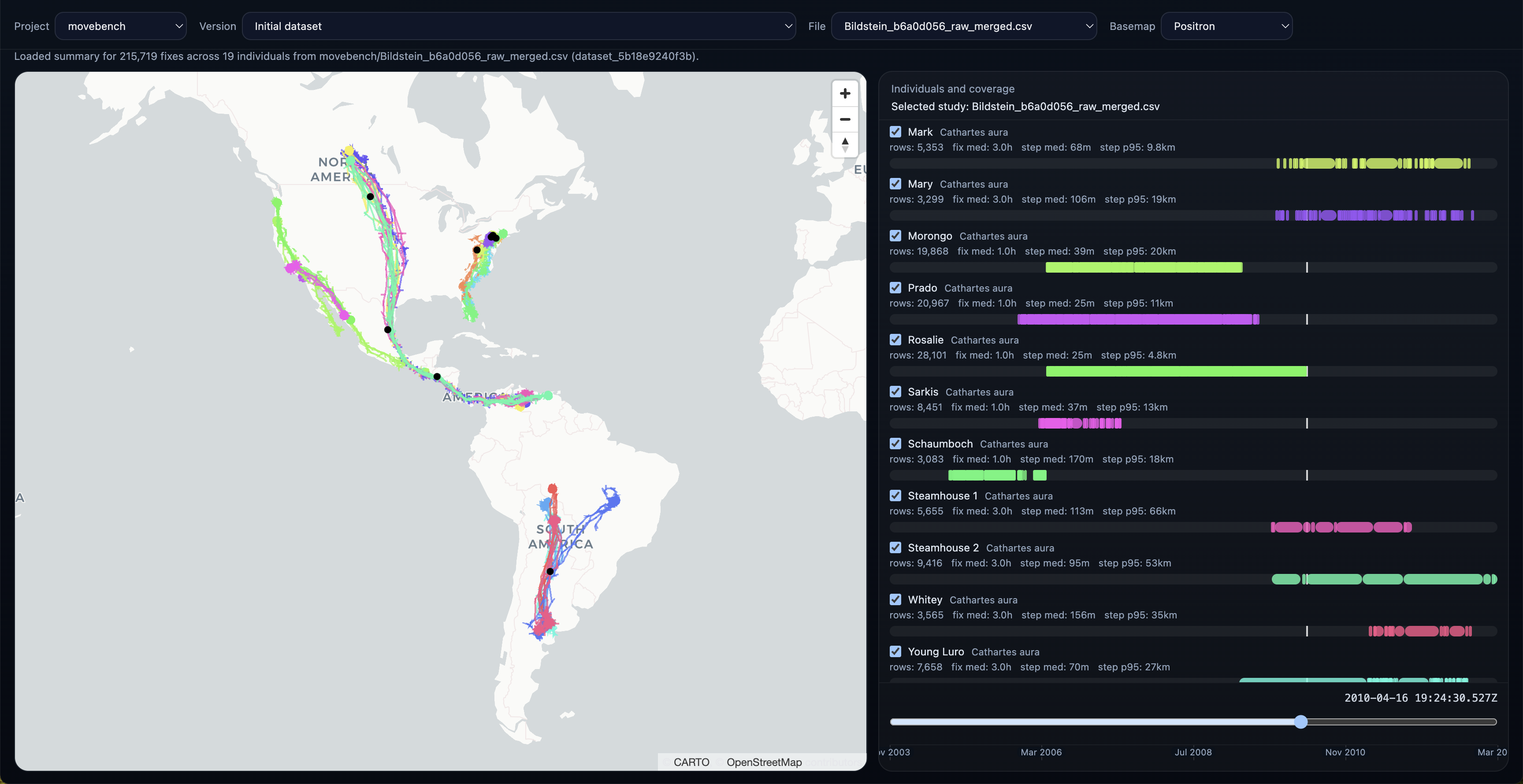}
    \caption{\textbf{Interactive data visualizer developed with \method{} for the \textsc{MoveBench} project.} Wildlife movement trajectories are displayed on an interactive map (left), with different colors representing the trajectories of different individuals. This is complemented by a timeline view (right) that displays the temporal coverage of each individual's trajectory and allows a user to navigate to different points in time using a slider bar (bottom right). The locations at the time indicated by the slider are highlighted by black dots on the map. This visualization tool was developed for data exploration, enabling us to identify data cleaning needs and verify any data transformations.}
    \label{fig:app}
\end{figure}

After acquiring the raw data, we developed a custom interface to visualize each study independently. We had in mind an interactive map with GPS tracks overlaid. We prompted \texttt{codex} to build this visualization in the frontend application, which it did in JavaScript and using deck.gl~\cite{deckgl}, Maplibre GL JS~\cite{maplibre_gl_js}, and CARTO basemaps~\cite{carto_basemaps}. We iterated on the design fairly extensively, adding features such as: different colored tracks for different individuals; a “current step” visualized as a black dot on each trajectory; an interactive timeline view with a slider to change the current step; checkboxes to show and hide different individuals; and a “preview” mode that reduces the number of points visualized in order to make loading studies faster. \cref{fig:app} shows the final state of the visualizer after this initial exploration.

Through this data visualization, we began to understand common patterns and gain a better intuition for the data as a whole. This also allowed us to identify issues we would need to address in order to clean and standardize the data for the benchmark. In particular, this initial visualization motivated discussions and decisions around key sources of variation and bias in the data:

\begin{enumerate}
    \item \textbf{Sample rates:} The time between GPS fixes differed substantially between studies, with some containing very high frequency recordings (\textit{e.g.} 1hz). We decided after conversations with domain experts that the shortest time horizon we would support is one hour, and that we would thin any high-frequency datasets accordingly.
    
    \item \textbf{Temporal extents:} Studies also differed in how long a sensor was deployed for, with some animals containing data for several years while others only contained data for a few days. We decided that in order to keep the benchmark to a tractable size we would need to cap the maximum number of points to include per individual.
    
    \item \textbf{Number of individuals:} Some studies included only a single individual, while others included hundreds. We decided that in order to maximize species diversity while keeping the benchmark size tractable, we would cap the maximum number of individuals included per study.
    
    \item \textbf{Subpopulations and leakage:} Some studies contained individuals from several subpopulations---\textit{i.e.}, groups of animals of the same species in different geographic locations. We decided that to improve the diversity of the benchmark we would aim to include individuals from each subpopulation. We also identified a potential source of leakage between training and test data: animals within each subpopulation often overlapped significantly in space and time. Therefore, when constructing train/test splits, we would need to ensure that any animals from the same subpopulation had the same cutoff timestamp between train and test data.
    
\end{enumerate}


These decisions motivated our next two phases of curation, an initial broad-strokes cleaning pass to address issues 1--3 (\cref{sec:terminalclean}), and interactive study-by-study cleaning to address issue 4 (\cref{sec:interactclean}).

\subsection{Initial terminal-based data cleaning}
\label{sec:terminalclean}

We performed initial broad-strokes cleaning operations in the terminal, within a \texttt{codex} instance launched inside the \method{} codebase. During this cleaning phase, the goal was to address issues 1--3 above.

First, to address issue 1, we implemented dataset-wide thinning---reducing the sampling rate to a maximum of one fix per hour, to support the benchmark's minimum forecast horizon discussed above.
As an illustrative example, the following shows a truncated (for space) version of the user-agent conversation for executing the thinning operation. The full conversation is included in Appendix~\cref{appendix:thinning}.

\vspace{5pt}
\begin{tcolorbox}[
    enhanced,
    breakable,
    colback=gray!4,
    colframe=gray!25,
    boxrule=0.5pt,
    arc=2.5mm,
    left=4mm,
    right=4mm,
    top=3mm,
    bottom=3mm,
    before skip=8pt,
    after skip=8pt
]

\textbf{\color{blue!70!black}User:}
The first thing I want to do is thin every study in \texttt{movebench} to
1 fix per hour per individual. Propose to me the algorithm you would use
to do this before starting.

\medskip

\textbf{\color{teal!70!black}Agent:}
I’m checking a few \texttt{movebench} file headers first so the thinning
algorithm matches the actual schema rather than guessing at column names
or timestamp format.

\smallskip

For \texttt{movebench}, I’d treat this as thinning every CSV in the current
dataset, independently per study file.

\medskip
Recommended algorithm:\\

\textit{..... See Appendix for full content .....}\\









One choice to confirm before I implement it: should ``1 fix per hour'' mean
this rolling 60-minute rule, or do you want one row per clock-hour bucket
instead?

\medskip

\textbf{\color{blue!70!black}User:}
Yes, I also want there to be some breathing room, say \(\pm 30\) minutes,
in case fixes are not exactly one hour apart.

\medskip

\textbf{\color{teal!70!black}Agent:}
I’d use an anchored hourly-thinning rule on every CSV in the current
\texttt{movebench} dataset.

\medskip
Algorithm:\\

\textit{..... See Appendix for full content .....}\\









One detail to confirm before I implement it: if \texttt{sensor-type} is
present, should I thin only \texttt{gps} rows, or all sensor rows that have
coordinates?

\medskip

\textbf{\color{blue!70!black}User:}
Yes, only GPS rows should remain. Make sure to do this as a DAG step as
outlined in the repo instructions.

\medskip

\textbf{\color{teal!70!black}Agent:}
To create the thinning as a persistent step, I need the attribution string
the app stores on steps. What user name should I record on the
\texttt{movebench} step?

\medskip

\textbf{\color{blue!70!black}User:}
\texttt{justin}
\end{tcolorbox}
\vspace{5pt}

Because the agent was launched inside the \method{} codebase and had the instruction scaffolding already loaded into context, it structured these operations as Python functions, bundled them into a single script, and submitted the script directly to the API. We provide the full execution record for reference in \cref{appendix:thinning}; at a high level the agent executed the following:

\vspace{5pt}
\newtcblisting{codeblock}{
    enhanced,
    breakable,
    listing only,
    listing engine=listings,
    colback=gray!4,
    colframe=gray!25,
    boxrule=0.5pt,
    arc=2.5mm,
    left=3mm,
    right=3mm,
    top=2mm,
    bottom=2mm,
    before skip=8pt,
    after skip=8pt,
    listing options={
        basicstyle=\ttfamily\footnotesize,
        breaklines=true,
        breakatwhitespace=false,
        columns=fullflexible,
        keepspaces=true,
        showstringspaces=false,
        tabsize=4,
        aboveskip=0pt,
        belowskip=0pt
    }
}

\begin{codeblock}
python - <<`PY'
from pathlib import Path
from app.execution import create_step

script = r"""
# Truncated for space (see Appendix).
# This string contains a full executable Python file, including imports, methods,
# and an if __name__ == `__main__' block.
"""

result = create_step(
    Path('data/movebench'),
    {
        `user': `justin',
        `title': (
            `Thin every study to one GPS fix per hour per '
            `individual (30-90 minute tolerance)'
        ),
        `operation': `filter_all_files',
        `kind': `python',
        `script': script,
    },
)
\end{codeblock}
\vspace{5pt}

Notice that the \texttt{script} variable contains a string representation of an entire Python file. On the backend, this is validated and then saved as a \texttt{.py} file, becoming the literal transformation stored in the DAG and used to create the next dataset version. 
While this may appear a counterintuitive design pattern, we have found that agents utilize it very effectively.

After this thinning, we were left with 14M points. Next, to address issue 2, 
we similarly prompted the agent to cap the maximum number of points per study. We decided to keep the equivalent of a maximum of two full years of data for each individual (a maximum of $17,520$ GPS points at 1-hour intervals), allowing the benchmark to capture any potential annually-varying movement patterns multiple times (anticipating the need for patterns to appear at least once in training data and once in testing data). Then, to address issue 3, we decided we would keep a maximum of six individuals per study total. Randomly sampling six would not work, since we had observed that some tracks had anomalies or were otherwise unusable. We prompted the agent to randomly sample fifteen, from which we decided we would select six by hand, a process we describe in \cref{sec:interactclean}.

Finally, we assigned preliminary train/test splits to each remaining individual in preparation to address issue 4. As discussed at the beginning of \cref{sec:casestudy}, we decided to support two test sets that would evaluate two different types of generalization: \textit{temporal} generalization---how well models trained on a set of individuals generalize to future movements of those same individuals---and \textit{individual} generalization---how well models trained on one set of individuals generalize to a different set of individuals of the same species. We decided we would pick held-out individuals either randomly or by hand during the subsequent interactive cleaning described in  \cref{sec:interactclean}. For held-out time points, we prompted the agent to create a preliminary 50/50 split, assigning the first 50\% of time points for each individual to the training set, and the second 50\% to the test set, with the plan to fine-tune these as well during interactive cleaning.

After these steps were executed, we inspected the datasets in the UI, where the following steps were visible in the DAG version dropdown menu:

\begin{figure}[h!]
    \centering
    \includegraphics[width=0.75\linewidth]{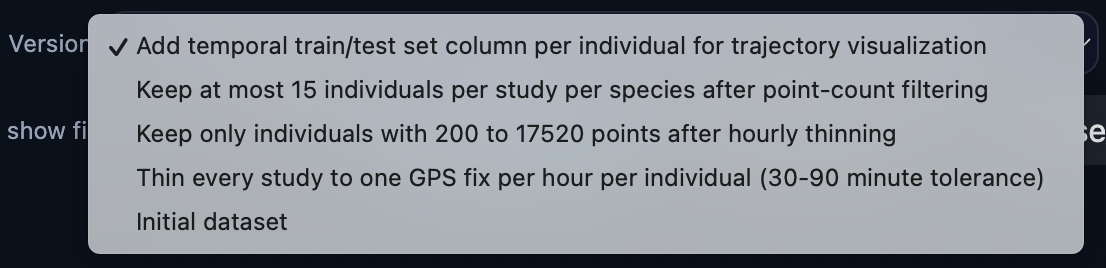}
    \label{fig:placeholder}
\end{figure}

With 5.9M data points remaining, we moved on to interactively curating the remaining samples.

\subsection{Iterative visualization and cleaning}
\label{sec:interactclean}

Now that initial curation was complete, the bulk of our effort followed the iterative loop visualized in \cref{fig:framework} (left): inspecting data through the visualization tools we had built, identifying additional data cleaning or visualization needs, and working in tandem with \texttt{codex} to implement them. 

For example, to address issue 3 from \cref{sec:initialviz}, we needed to choose which six of the fifteen remaining individuals from each study to keep for the benchmark. We went study-by-study, inspecting individuals for anomalies or any other reasons to remove them, and then randomly selecting six of the remaining group per-study to keep. We initially performed these operations---deleting anomalous individuals, and randomly selecting the remaining six---by prompting \texttt{codex} through the terminal, but quickly realized it would be faster to have a button in the frontend application to do this automatically. We prompted \texttt{codex} to add this functionality. We also implemented a pop-up confirmation for every transformation step submitted through the UI that summarized the change and included two text fields: ``username’’ and ``reason’’. The latter allowed us to record why we made each change.

In order to comply with the API requirements, similarly to the terminal-based Python execution in \cref{sec:terminalclean}, \texttt{codex} implemented these as Python functions that themselves return a string representation of another Python file containing transformation code, \textit{e.g.} (truncated here for space but included in full in Appendix~\cref{appendix:python}):

\vspace{5pt}
\newtcblisting{codeblock2}{
    enhanced,
    breakable,
    listing only,
    listing engine=listings,
    colback=gray!4,
    colframe=gray!25,
    boxrule=0.5pt,
    arc=2.5mm,
    left=3mm,
    right=3mm,
    top=2mm,
    bottom=2mm,
    before skip=8pt,
    after skip=8pt,
    listing options={
        basicstyle=\ttfamily\footnotesize,
        breaklines=true,
        breakatwhitespace=false,
        columns=fullflexible,
        keepspaces=true,
        showstringspaces=false,
        tabsize=4,
        aboveskip=0pt,
        belowskip=0pt
    }
}

\begin{codeblock2}
def build_random_subsample_individuals_script(keep_individuals: list[str], selected_individuals: list[str], reason: str) -> str:

    return f"""import csv
               import json
               import os
               from pathlib import Path

               KEEP = set({json.dumps(keep_individuals)})
               SELECTED = set({json.dumps(selected_individuals)})
               REASON = {json.dumps(reason)}

               def main():
                   # Write a new data file containing only individuals in KEEP
                   # (See Appendix for full code)

               if __name__ == "__main__":
                   main()
    """
\end{codeblock2}
\vspace{5pt}

Throughout the rest of the cleaning process, we implemented additional common operations that continued to follow this format. These included: keep or delete all selected individuals; align the timestamp of the train/test split for each selected individual (to address issue 4 from \cref{sec:initialviz}); assign selected individuals to the held-out individual test set; remove a study entirely (if it contained no usable data points); undo the last operation; and mark the current study complete. We modified the frontend to sort studies by their completion status, so that every time we opened the application we would be met with an incomplete study to begin work on.

\subsection{Final results}
Starting with 89M raw GPS coordinates across 234 studies, we used the \method{} framework to create \textsc{MoveBench}: a benchmark dataset of 2.6M GPS coordinates curated from 807 individuals from 110 different species, sourced from 147 independent studies. The curation effort took a team of two curators a total of three days to complete. The data provenance history, including all curation decisions, will be published alongside the benchmark to enable reproducibility from raw data to finished dataset.

\section{Discussion and limitations}

We introduce \method{}, a framework that makes agentic data curation more reproducible and verifiable via data provenance and a human-in-the-loop visual validation workflow. We have found the framework to be very useful, and we use it daily in our own data work, including the development of the \textsc{MoveBench} benchmark. We hope that others will also find it useful and build upon it to enable their own bespoke data curation and/or cleaning workflows. The framework is not limited to the use cases discussed in this manuscript; it is meant to be a flexible starting point for any number of data-centric applications, and we look forward to seeing what others will build with it.


There are limitations, risks, and cost/value tradeoffs to consider when using any LLM-based workflows, including the \method{} framework. 
In particular, while the framework provides instructions for guiding agent actions, it cannot provide absolute \textit{guarantees} against unwanted actions---for instance, bypassing the application's intended use and modifying data directly anyway. While we have not experienced this failure mode ourselves, 
we recommend taking addition steps to keep data safe while experimenting with the tool---for instance, utilizing this workflow only on data that has been securely backed up, and potentially removing the agent from the loop once a curation process is verified and fully implemented in the UI.
Further, the use of proprietary LLMs comes with financial costs, data privacy concerns, and environmental resource use that must be evaluated in the context of the goals of any particular data curation project. 

If carefully considered and designed, agentic coding tools can enable experts to more quickly build the tooling they need to move from raw data collections to scientific analysis. We hope the \method{} framework will provide a useful foundation for doing so.

\section*{Acknowledgments}

Thanks to members of the Beery Lab at MIT and the MoveBON Working Group on AI for helpful feedback and hands-on testing. JK and SaB were supported by NSF award 2330423 and NSERC award 585136. JK was additionally supported by NSF award 2313998. ShB was supported by a CHE Fellowship for Marine Sciences.

\clearpage

\bibliographystyle{plain}
\bibliography{bib}

\clearpage

\appendix
\section*{Appendix}
\label{appendix:api}

\section{API specification}
\label{appendix:api}

Out of the box, we provide the following API endpoints:

\begin{itemize}
    \item \texttt{GET /api/projects}: list available projects.
    \item \texttt{GET /api/project/\{project\}/state}: retrieve the current state of a project, including the current dataset and recorded history.
    \item \texttt{GET /api/project/\{project\}/graph}: retrieve the provenance graph, including dataset nodes and cleaning-step edges.
    \item \texttt{GET /api/project/\{project\}/dataset/\{dataset\_id\}}: retrieve a manifest for a particular dataset version, containing the dataset identifier and a list of data files included in that version.
    \item \texttt{GET /api/project/\{project\}/artifact/\{dataset\_id\}/\{logical\_name\}}: retrieve a data file from a dataset version.
    \item \texttt{GET /api/project/\{project\}/artifact/\{dataset\_id\}/\{logical\_name\}/meta}: retrieve metadata for a data file.
    \item \texttt{GET /api/project/\{project\}/artifact/\{dataset\_id\}/\{logical\_name\}/preview}: retrieve a lightweight preview of a data file.
    \item \texttt{POST /api/project/\{project\}/analyses}: create an exploratory analysis that is recorded but does not create a new dataset version.
    \item \texttt{POST /api/project/\{project\}/steps}: create a persistent cleaning step that produces a new dataset version.
    \item \texttt{POST /api/project/\{project\}/head}: move the current-dataset pointer to an existing dataset version.
    \item \texttt{POST /api/project/\{project\}/undo}: move the current-dataset pointer to the parent of the current dataset.
\end{itemize}

\section{Additional DAG implementation details}
\label{appendix:dag}

\textbf{DAG nodes.} To avoid duplicating the entire collection of data files in every node, we represent each node as a manifest describing the current collection of files in that version.
Any data files that change between dataset versions are copied to the child node, while unaffected files are reused by reference rather than copied. 
Previous versions of the dataset can be accessed without deleting any nodes or edges from the graph by moving this pointer to another existing dataset version.  Moving this pointer backward can also be used to \textit{undo} operations without deleting the records of the undone step(s).
If a new cleaning step is created after the current-dataset pointer has been moved to an earlier node, the resulting child node forms a new branch in the provenance graph. Existing descendant nodes remain accessible and may be restored by moving the pointer again.

\textbf{DAG edges.}
DAG edges store a record of each transformation step. This record includes the executable script that performs the transformation as well as JSON metadata files that log the original user request, the identity of the requesting user, and a list of affected data files.
If a transformation script exits with an error, the framework does not create an edge or child node in the provenance graph.

\textbf{Exploratory analyses. }
The framework also distinguishes between two kinds of data operations performed within a project. The primary operation is a \emph{step}, which involves modifying the dataset (\textit{i.e.} creating a new dataset version). However, the choice of whether or not to modify a dataset may require some initial analyses. AI agents perform notably well at data analysis, often by writing and executing standalone scripts and inspecting outputs before returning a summary of takeaways to the user. To support this, we also enable \emph{analysis} operations that do not modify the dataset. 
Analyses may create and execute scripts, which are stored on disk for reference, but unlike steps they do not create new nodes in the provenance graph. 



\section{Agent conversation examples}



\subsection{Dataset thinning}
\label{appendix:thinning}

The full conversation leading to the hourly thinning of trajectory data for \textsc{MoveBench} was:

\begin{tcolorbox}[
    enhanced,
    breakable,
    colback=gray!4,
    colframe=gray!25,
    boxrule=0.5pt,
    arc=2.5mm,
    left=4mm,
    right=4mm,
    top=3mm,
    bottom=3mm,
    before skip=8pt,
    after skip=8pt
]

\textbf{\color{blue!70!black}User:}
The first thing I want to do is thin every study in \texttt{movebench} to
1 fix per hour per individual. Propose to me the algorithm you would use
to do this before starting.

\medskip

\textbf{\color{teal!70!black}Agent:}
I’m checking a few \texttt{movebench} file headers first so the thinning
algorithm matches the actual schema rather than guessing at column names
or timestamp format.

\smallskip

For \texttt{movebench}, I’d treat this as thinning every CSV in the current
dataset, independently per study file.

\medskip
Recommended algorithm:

\begin{enumerate}
    \item Parse each row as a candidate fix only if it has
    \texttt{timestamp}, \texttt{individual-local-identifier}, and valid
    coordinates (\texttt{location-long}, \texttt{location-lat}). If
    \texttt{sensor-type} exists, keep only location fixes such as
    \texttt{gps}.

    \item Within each file, group rows by
    \texttt{individual-local-identifier}.

    \item For each individual, sort by \texttt{timestamp}, then break exact
    ties with \texttt{event-id} if present.

    \item Keep the first valid fix, then keep the next fix only when it is
    at least 60 minutes after the last kept fix.

    \item Drop the rest.
\end{enumerate}

I recommend this over simple ``calendar hour bins'' because hour bins can
still keep two points a minute apart across an HH:59 / HH+1:00 boundary.
The rolling 60-minute rule guarantees the retained fixes are actually
spaced by at least one hour.

\medskip
Deterministic tie rules:

\begin{itemize}
    \item If multiple rows share the same timestamp, keep the one with the
    smallest \texttt{event-id}.
    \item If \texttt{event-id} is missing, fall back to file order.
\end{itemize}

One choice to confirm before I implement it: should ``1 fix per hour'' mean
this rolling 60-minute rule, or do you want one row per clock-hour bucket
instead?

\medskip

\textbf{\color{blue!70!black}User:}
Yes, I also want there to be some breathing room, say \(\pm 30\) minutes,
in case fixes are not exactly one hour apart.

\medskip

\textbf{\color{teal!70!black}Agent:}
I’d use an anchored hourly-thinning rule on every CSV in the current
\texttt{movebench} dataset.

\medskip
Algorithm:

\begin{enumerate}
    \item In each file, keep only rows with a parseable
    \texttt{timestamp}, an \texttt{individual-local-identifier}, and valid
    coordinates.

    \item Group by \texttt{individual-local-identifier} and sort each group
    by \texttt{timestamp}, then \texttt{event-id} as a deterministic
    tie-breaker.

    \item Keep the first fix for that individual as the anchor.

    \item For the next retained fix, look for candidates relative to the
    last kept time:
    \begin{itemize}
        \item Ignore anything less than 30 minutes later.
        \item Among fixes 30--90 minutes later, keep exactly one: the fix
        closest to 60 minutes after the last kept fix.
        \item If there is a tie, keep the earlier timestamp, then the
        smaller \texttt{event-id}.
    \end{itemize}

    \item After keeping that fix, repeat from the new anchor.

    \item If there is no fix in the 30--90 minute window and the next
    available fix is more than 90 minutes later, keep that next fix and
    treat it as a gap in coverage rather than trying to invent an hourly
    point.
\end{enumerate}

Why I’d use this:

\begin{itemize}
    \item It enforces the ``about one per hour'' goal without requiring
    exact 60-minute spacing.
    \item It avoids keeping multiple fixes inside the same approximate hour.
    \item It behaves sensibly when a tag misses one or more expected fixes.
\end{itemize}

One detail to confirm before I implement it: if \texttt{sensor-type} is
present, should I thin only \texttt{gps} rows, or all sensor rows that have
coordinates?

\medskip

\textbf{\color{blue!70!black}User:}
Yes, only GPS rows should remain. Make sure to do this as a DAG step as
outlined in the repo instructions.

\medskip

\textbf{\color{teal!70!black}Agent:}
To create the thinning as a persistent step, I need the attribution string
the app stores on steps. What user name should I record on the
\texttt{movebench} step?

\medskip

\textbf{\color{blue!70!black}User:}
\texttt{justin}

\end{tcolorbox}

The full script submitted was:


\begin{codeblock}
python - <<'PY'
from pathlib import Path
from app.execution import create_step

script = r"""
import csv
import json
import os
from datetime import timedelta
from pathlib import Path

import duckdb

TIME_FORMATS = [
    '%Y-%m-%d %H:%M:%S.%f',
    '%Y-%m-%d %H:%M:%S',
    '%Y-%m-%dT%H:%M:%S.%f',
    '%Y-%m-%dT%H:%M:%S',
    '%Y-%m-%dT%H:%M:%S.%fZ',
    '%Y-%m-%dT%H:%M:%SZ',
]
BATCH_SIZE = 50000
MAX_BIGINT = 9223372036854775807


def q(name: str) -> str:
    return '"' + name.replace('"', '""') + '"'


def sql_quote(value: str) -> str:
    return "'" + value.replace("'", "''") + "'"


def build_sorted_query(path: Path, header: list[str]) -> str:
    raw_cols = ', '.join(f't.{q(col)}' for col in header)

    timestamp_expr = 'COALESCE(' + ', '.join(
        f"try_strptime(t.{q('timestamp')}, '{fmt}')"
        for fmt in TIME_FORMATS
    ) + ')'

    sensor_candidates = []
    for col in ('sensor-type', 'sensor-type-id'):
        if col in header:
            sensor_candidates.append(
                f"NULLIF(TRIM(t.{q(col)}), '')"
            )

    sensor_pred = (
        f"lower(COALESCE({', '.join(sensor_candidates)})) = 'gps'"
        if sensor_candidates
        else 'FALSE'
    )

    event_expr = (
        f"COALESCE("
        f"TRY_CAST(t.{q('event-id')} AS BIGINT), "
        f"{MAX_BIGINT})"
        if 'event-id' in header
        else str(MAX_BIGINT)
    )

    return f'''
        SELECT
            {raw_cols},
            __parsed_ts,
            __event_sort,
            __source_order
        FROM (
            SELECT
                {raw_cols},
                {timestamp_expr} AS __parsed_ts,
                {event_expr} AS __event_sort,
                row_number() OVER () AS __source_order
            FROM read_csv_auto(
                {sql_quote(str(path))},
                header=true,
                all_varchar=true
            ) AS t
        ) t
        WHERE {sensor_pred}
          AND NULLIF(
              TRIM(t.{q('individual-local-identifier')}),
              ''
          ) IS NOT NULL
          AND TRY_CAST(
              t.{q('location-long')} AS DOUBLE
          ) IS NOT NULL
          AND TRY_CAST(
              t.{q('location-lat')} AS DOUBLE
          ) IS NOT NULL
          AND __parsed_ts IS NOT NULL
        ORDER BY
            t.{q('individual-local-identifier')},
            __parsed_ts,
            __event_sort,
            __source_order
    '''


def choose_best(candidates, target):
    return min(
        candidates,
        key=lambda row: (
            abs((row[-3] - target).total_seconds()),
            row[-3],
            row[-2],
            row[-1],
        ),
    )


def write_row(writer, row, raw_len):
    writer.writerow(row[:raw_len])


def thin_file(input_path: Path, output_path: Path) -> dict:
    with input_path.open(
        'r',
        newline='',
        encoding='utf-8'
    ) as handle:
        header = next(csv.reader(handle))

    individual_idx = header.index(
        'individual-local-identifier'
    )
    raw_len = len(header)

    stats = {
        'logical_name': input_path.name,
        'eligible_rows': 0,
        'retained_rows': 0,
    }

    con = duckdb.connect()

    current_individual = None
    last_kept_ts = None
    window_candidates = []

    def keep_candidate(writer):
        nonlocal last_kept_ts, window_candidates

        if not window_candidates:
            return False

        chosen = choose_best(
            window_candidates,
            last_kept_ts + timedelta(hours=1)
        )

        write_row(writer, chosen, raw_len)
        stats['retained_rows'] += 1
        last_kept_ts = chosen[-3]
        window_candidates = []

        return True

    try:
        query = build_sorted_query(input_path, header)
        cursor = con.execute(query)

        with output_path.open(
            'w',
            newline='',
            encoding='utf-8'
        ) as handle:
            writer = csv.writer(handle)
            writer.writerow(header)

            while True:
                rows = cursor.fetchmany(BATCH_SIZE)

                if not rows:
                    break

                stats['eligible_rows'] += len(rows)
                idx = 0

                while idx < len(rows):
                    row = rows[idx]
                    ts = row[-3]
                    individual = row[individual_idx]

                    if individual != current_individual:
                        keep_candidate(writer)
                        current_individual = individual
                        last_kept_ts = ts
                        window_candidates = []

                        write_row(writer, row, raw_len)
                        stats['retained_rows'] += 1

                        idx += 1
                        continue

                    window_start = (
                        last_kept_ts
                        + timedelta(minutes=30)
                    )
                    window_end = (
                        last_kept_ts
                        + timedelta(minutes=90)
                    )

                    if ts < window_start:
                        idx += 1
                        continue

                    if ts <= window_end:
                        window_candidates.append(row)
                        idx += 1
                        continue

                    if keep_candidate(writer):
                        continue

                    write_row(writer, row, raw_len)
                    stats['retained_rows'] += 1
                    last_kept_ts = ts
                    idx += 1

            keep_candidate(writer)

    finally:
        con.close()

    return stats


def main():
    spec_path = Path(
        os.environ['CLAUDAG_SPEC_PATH']
    )
    summary_path = Path(
        os.environ['CLAUDAG_SUMMARY_PATH']
    )

    spec = json.loads(spec_path.read_text())

    file_summaries = []
    total_eligible = 0
    total_retained = 0
    zero_retained = []

    for input_spec, output_spec in zip(
        spec['input_files'],
        spec['output_files']
    ):
        result = thin_file(
            Path(input_spec['path']),
            Path(output_spec['path'])
        )

        file_summaries.append(result)
        total_eligible += result['eligible_rows']
        total_retained += result['retained_rows']

        if result['retained_rows'] == 0:
            zero_retained.append(
                result['logical_name']
            )

    summary = {
        'operation': spec['operation'],
        'rule': (
            'Keep GPS rows with valid coordinates and timestamps; '
            'per individual keep the first fix, then the fix closest '
            'to +60 minutes within a 30-90 minute window; if no '
            'candidate appears within 30-90 minutes, keep the next '
            'later fix as a gap reset.'
        ),
        'files_processed': len(file_summaries),
        'eligible_rows': total_eligible,
        'retained_rows': total_retained,
        'files_with_zero_retained_rows': zero_retained,
        'file_summaries': file_summaries,
    }

    summary_path.write_text(
        json.dumps(
            summary,
            indent=2,
            sort_keys=True
        ) + '\n'
    )

    print(json.dumps({
        'files_processed': len(file_summaries),
        'retained_rows': total_retained,
        'zero_retained': len(zero_retained),
    }))


if __name__ == '__main__':
    main()
"""

result = create_step(
    Path('data/movebench'),
    {
        'user': 'justin',
        'title': (
            'Thin every study to one GPS fix per hour per '
            'individual (30-90 minute tolerance)'
        ),
        'operation': 'filter_all_files',
        'kind': 'python',
        'script': script,
    },
)

print(result['step']['step_id'])
print(result['step']['output_dataset_id'])
print(result['step']['summary'].get('retained_rows'))
PY
\end{codeblock}

\subsection{Sample Python function submitted to the API}
\label{appendix:python}


\begin{codeblock}

def build_random_subsample_individuals_script(keep_individuals: list[str], selected_individuals: list[str], reason: str) -> str:
    return f"""import csv
import json
import os
from pathlib import Path

KEEP = set({json.dumps(keep_individuals)})
SELECTED = set({json.dumps(selected_individuals)})
REASON = {json.dumps(reason)}


def main():
    spec = json.loads(Path(os.environ["CLAUDAG_SPEC_PATH"]).read_text())
    input_path = Path(spec["input_files"][0]["path"])
    output_path = Path(spec["output_files"][0]["path"])
    temp_path = output_path.with_suffix(output_path.suffix + ".tmp")

    seen_selected = set()
    kept_individuals = set()
    removed_rows = 0
    retained_rows = 0
    removed_rows_by_individual = {{}}

    with input_path.open("r", newline="", encoding="utf-8") as src, temp_path.open("w", newline="", encoding="utf-8") as dst:
        reader = csv.DictReader(src)
        fieldnames = list(reader.fieldnames or [])
        writer = csv.DictWriter(dst, fieldnames=fieldnames)
        writer.writeheader()
        for row in reader:
            individual = (row.get("individual-local-identifier") or "").strip()
            if individual in SELECTED:
                seen_selected.add(individual)
                if individual not in KEEP:
                    removed_rows += 1
                    removed_rows_by_individual[individual] = removed_rows_by_individual.get(individual, 0) + 1
                    continue
                kept_individuals.add(individual)
            writer.writerow(row)
            retained_rows += 1

    missing = sorted(SELECTED - seen_selected)
    if missing:
        temp_path.unlink(missing_ok=True)
        raise SystemExit("Selected individuals were not found in the file: " + ", ".join(missing))
    if retained_rows == 0:
        temp_path.unlink(missing_ok=True)
        raise SystemExit("Random subsampling would leave the file empty. Remove the whole study instead.")

    temp_path.replace(output_path)

    summary = {{
        "operation": "random_subsample_individuals",
        "reason": REASON,
        "selected_individuals": sorted(SELECTED),
        "kept_individuals": sorted(kept_individuals),
        "removed_individuals": sorted(SELECTED - KEEP),
        "requested_sample_size": len(KEEP),
        "removed_rows": removed_rows,
        "retained_rows": retained_rows,
        "removed_rows_by_individual": removed_rows_by_individual,
    }}
    Path(os.environ["CLAUDAG_SUMMARY_PATH"]).write_text(json.dumps(summary, indent=2, sort_keys=True) + "\\n")


if __name__ == "__main__":
    main()
"""

\end{codeblock}

\end{document}